The susceptibility tensor of a uniform magnetized plasma with a transverse electric field and the electromagnetic waves propagating parallel to the magnetic field

Deng Zhou[1], Yue Ming[2],   and Jinfang Wang[1]

[1]Institute of Plasma Physics, Chinese Academy of Sciences, Hefei 230031, P. R. China

[2]School of Opo-electronic Information Science and Technology, Yantai University, Yantai 264005, P. R. China

Abstract

In natural and laboratory magnetized plasmas, an equilibrium electric field may exist perpendicular to the background magnetic field. In such a situation all the plasma species experience a common drift and the unperturbed distribution functions have a common shift in velocity space.   In this work the susceptibility tensor is first derived in the laboratory frame for such a situation using the commonly used method of integration along unperturbed trajectory. Then the Lorentz transformation method is adopted to verify the results. As an application we give an analysis of the waves propagating parallel to the background magnetic field for a simple electron plus single charged ion plasma. A qualitative new phenomenon is the appearance of a resonance at the plasma Langmuir frequency if a transverse equilibrium electric field is present.

I.       INTRODUCTION

The kinetic theory of electromagnetic waves for plasmas in a static uniform magnetic field was developed in 1950s by different researchers. The standard method and formulas can be found nowadays in textbooks and monographs[1,2]. In some circumstances, an equilibrium electric field perpendicular to the background magnetic field may be present, e. g. near the edge region of magnetic confinement devices[3-6], where a transport barrier may be induced due to the radial electric field. In future larger devices like the International Thermonuclear Experimental Reactor (ITER), the electric field my be present in a wider region near the edge due to its scaling with the major radius. The effect of the radial electric field in tokamaks has been under investigation for many years. The radio frequency waves have long been used in plasma heating and current drive in magnetic confinement devices[7]. People made use of the ray-tracing as well the full wave approaches to simulate the radio frequency waves in tokamak plasmas[8,9].   However, so far no basic theory on radio frequency waves in magnetized plasmas has considered the effect of equilibrium electric fields, not to say the application of waves.

In this work, we derive the susceptibility tensor for linear response of a homogenous plasma to the electromagnetic waves. The plasma is immersed in a uniform magnetic field with a uniform transverse electric field present. Under such a situation, all plasma species experience an $\mathbf{E} \times \mathbf{B}$ drift, so the unperturbed particle has distribution functions with a common shift velocity in the laboratory frame and the unperturbed particle trajectory has a transverse drift velocity in addition to the normal cyclotron motion. We take the method of integration along unperturbed trajectory to obtain the perturbed distribution function, which was first developed by Drummond[10]. We follow the procedure closely similar to that given by Stix[1]. The process is complicated but straightforward. We first derive the general form of the susceptibility tensor expressed by the integrals in velocity space, then derive the form for the shifted Maxwellian perpendicular distribution. Then we adopt the Lorentz transformation method to verify the results.

As an example of application we give an analysis of the waves propagating parallel to the background magnetic field for a simple electron plus single charged ion plasma. In the absence of the equilibrium electric field, there are three independent normal modes, two circularly polarized electromagnetic waves and one longitudinal electrostatic wave. The three modes are coupled in the presence of an electric field and their polarizations are also changed.

In the following we will present the derivation of the susceptibility tensor using the method of integration along unperturbed trajectory in Sec. II and test the results using the Lorentz transformation method in Sec. III. The analysis of parallel propagating waves is presented in Sec. IV. The final section is devoted to the summary and the discussion.

## II.    DERIVATION OF THE SUSCEPTIBILITY TENSOR

In this section we derive the general form of the susceptibility tensor for a plasma immersed in a uniform magnetic field with a transverse uniform electric field. We closely follow the approaches presented in Stix's book[1]. We use the cgs units in the derivation.

In the Cartesian coordinate system $(\hat{\mathbf{x}}, \hat{\mathbf{y}}, \hat{\mathbf{z}})$, a uniform background magnetic field is directed along $\hat{\mathbf{z}}$ given by $\mathbf{B}_0 = B_0 \hat{\mathbf{z}}$, and a uniform background electric field is directed along $\hat{\mathbf{x}}$ given by $\mathbf{E}_0 = -E_0 \hat{\mathbf{x}}$. Then, the unperturbed Vlasov equation is

$$\mathbf{v} \cdot \nabla f_0 + \left( \mathbf{E}_0 + \frac{1}{c} \mathbf{v} \times \mathbf{B}_0 \right) \cdot \nabla_{\mathbf{v}} f_0 = 0 \qquad (1)$$

For a spacially uniform plasma, the first term is not involved. It is ready to solve Eq. (1) to get unperturbed distribution function in the form

$$f_0(\mathbf{v}) = f_0(v_z, v_\perp) \quad (2)$$

Where $v_z = \mathbf{v} \cdot \hat{\mathbf{z}}$ is the parallel velocity of the particle along the magnetic field and

$$v_\perp = \sqrt{v_x^2 + \left( v_y - v_d \right)^2} \qquad (3)$$

With $v_d = \dfrac{cE_0}{B_0}$ the drift of particles along $\hat{\mathbf{y}}$ caused by the electric field. $v_d$ is used to represent the strength of electric field in the following. We adopt the method of integration along unperturbed trajectory to obtain the perturbed distribution function. The unperturbed trajectory obeys the equation

$$\frac{d\mathbf{v}'}{dt'} = \frac{q}{m} \left( \mathbf{E}_0 + \frac{1}{c} \mathbf{v}' \times \mathbf{B}_0 \right) \qquad (4)$$

A prime denotes the quantity at the varying time $t'$. The solution to Eq. (4) is given by

$$\mathbf{v}' = v_x'\hat{\mathbf{x}} + v_y'\hat{\mathbf{y}} + v_z'\hat{\mathbf{z}} \quad (5a)$$

$$v_x'(t') = v_\perp \cos(\phi + \Omega\tau) \quad (5b)$$

$$v_y'(t') = v_d + v_\perp \sin(\phi + \Omega\tau) \quad (5c)$$

$$v_z'(t') = v_z \quad (5d)$$

Where $\tau = t - t'$, and $\Omega = \dfrac{qB_0}{mc}$ the gyro-frequency. In Eqs. (5) $v_z$ and $v_\perp$ are invariables of particle motion. At time $t' = t$, the velocity is $v_x = v_\perp \cos(\phi)$ and $v_y = v_d + v_\perp \sin(\phi)$.
The trajectory in configuration space is then given by

$$x' = x - \frac{v_\perp}{\Omega}\left[\sin(\phi + \Omega\tau) - \sin\phi\right] \quad (6a)$$

$$y' = y + \frac{v_\perp}{\Omega}\left[\cos(\phi + \Omega\tau) - \cos\phi\right] - v_d\tau \quad (6b)$$

$$z' = z - v_z\tau \quad (6c)$$

We write the perturbed electric field $\mathbf{E}_1(\mathbf{r}',t') = \mathbf{E}\exp\left[-i(\omega t' - \mathbf{k}\cdot\mathbf{r}')\right]$ and decompose the wave vector $\mathbf{k} = k_\perp\cos\theta\hat{\mathbf{x}} + k_\perp\sin\theta\hat{\mathbf{y}} + k_z\hat{\mathbf{z}}$. From the perturbed Vlasov equation, using

Faraday's induction law to relate the perturbed magnetic field to the electric field, one can

readily obtain the perturbed distribution as

$$f_1(\mathbf{r},\mathbf{v},t) = -\frac{q}{m}\exp\left[-i(\omega t - \mathbf{k}\cdot\mathbf{r})\right]$$

$$\int_0^{+\infty} d\tau \frac{e^{i\beta}}{\omega}\{U\cos(\phi + \Omega\tau)E_x + U\sin(\phi + \Omega\tau)E_y + DE_y + VE_z\} \quad (7)$$

Here we have introduced symbols

$$U = \tilde{\omega}f_{0\perp} + k_z\left(v_\perp f_{0z} - v_z f_{0\perp}\right) \quad (8a)$$

$$D = v_d\left[\cos(\phi - \theta + \Omega\tau)k_\perp f_{0\perp} + k_z f_{0z}\right] \quad (8b)$$

$$V = \tilde{\omega}f_{0z} - k_\perp\left(v_\perp f_{0z} - v_z f_{0\perp}\right)\cos(\phi - \theta + \Omega\tau) \quad (8c)$$

$$\beta = -z\left[\sin(\phi - \theta + \Omega\tau) - \sin(\phi - \theta)\right] + (\tilde{\omega} - k_z v_z)\tau \quad (8d)$$

with $z = \dfrac{k_\perp v_\perp}{\Omega}$, $\tilde{\omega} = \omega - k_y v_d$, $f_{0\perp} = \dfrac{\partial f_0(v_z,v_\perp)}{\partial v_\perp}$ and $f_{0z} = \dfrac{\partial f_0(v_z,v_\perp)}{\partial v_z}$.

To carry out the integration in Eq. (7), we use the well known identity

$$e^{iz\sin\phi} = \sum_{n=-\infty}^{+\infty} J_n(z) e^{in\phi} \quad \text{to reach}$$

$$e^{i\beta} = \sum_{m,n=-\infty}^{+\infty} J_m(z) J_n(z) e^{i(m-n)(\phi-\theta)} e^{i(\tilde{\omega}-k_z v_z - n\Omega)} \qquad (9)$$

Where $J_n(z)$ is the n-th order Bessel function. Substitution of (9) in (7), a straightforward calculation yields

$$f_1(\mathbf{r},\mathbf{v},t) = -i\frac{q}{m\omega} \sum_{m,n=-\infty}^{+\infty} \frac{J_n e^{i(m-n)(\phi-\theta)}}{\tilde{\omega}-k_z v_z - n\Omega}$$

$$\left[ \left( \frac{nJ_n}{z}\cos\theta - iJ_n'\sin\theta \right) U E_x + \left( \frac{nJ_n}{z}\sin\theta + iJ_n'\cos\theta \right) U E_y + J_n v_d X E_y + J_n W E_z \right] \qquad (10)$$

with

$$X = \frac{n\Omega}{v_\perp} f_{0\perp} + k_z f_{0z} \qquad (11a)$$

$$W = (\tilde{\omega}-n\Omega) f_{0z} + \frac{n\Omega}{v_\perp} v_z f_{0\perp} \qquad (11b)$$

In (10), we have dropped the common factor $\exp\left[-i(\omega t - \mathbf{k}\cdot\mathbf{r})\right]$ for conciseness. The current perturbation is $j_1 = \sigma \cdot \mathbf{E}_1 = \int q\mathbf{v} f_1(\mathbf{r},\mathbf{v},t) d^3\mathbf{v}$ and the susceptibility tensor is related to the conductivity tensor through $\boldsymbol{\chi} = \frac{4\pi i}{\omega} \boldsymbol{\sigma}$. For the equilibrium distribution independent of gyro-angle $\phi$, noting that $\mathbf{v} = v_\perp \cos\phi\hat{\mathbf{x}} + (v_d + v_\perp\sin\phi)\hat{\mathbf{y}} + v_z\hat{\mathbf{z}}$ and $d^3\mathbf{v} = v_\perp d\phi\, dv_\perp dv_z$, one can readily perform the integration with respect to $\phi$ in evaluating the current and get the susceptibility tensor contributed by species $s$

$$\tilde{\boldsymbol{\chi}}_s = \frac{\omega_{ps}^2}{\omega^2} \int_0^{+\infty} 2\pi v_\perp dv_\perp \int_{-\infty}^{+\infty} dv_z \left[ \hat{\mathbf{z}}\hat{\mathbf{z}} \left( \frac{f_{0z}}{v_z} - \frac{f_{0\perp}}{v_\perp} \right) v_z^2 + \sum_{n=-\infty}^{+\infty} \frac{v_\perp U \tilde{\mathbf{T}}_n + v_d \tilde{\mathbf{R}}_n}{\tilde{\omega}-k_z v_z - n\Omega_s} \right] \qquad (12)$$

Where the matrices

$$\tilde{\mathbf{T}}_n = \begin{bmatrix} \left(\frac{nJ_n}{z}\cos\theta\right)^2 + \left(J_n'\sin\theta\right)^2 & \frac{inJ_nJ_n'}{z} + J_{n-1}J_{n+1}\sin\theta\cos\theta & J_n\left(\frac{nJ_n}{z}\cos\theta + iJ_n'\sin\theta\right)\frac{v_z}{v_\perp} \\ -\frac{inJ_nJ_n'}{z} + J_{n-1}J_{n+1}\sin\theta\cos\theta & \left(J_n'\cos\theta\right)^2 + \left(\frac{nJ_n}{z}\sin\theta\right)^2 & J_n\left(\frac{nJ_n}{z}\sin\theta - iJ_n'\cos\theta\right)\frac{v_z}{v_\perp} \\ J_n\left(\frac{nJ_n}{z}\cos\theta - iJ_n'\sin\theta\right)\frac{v_z}{v_\perp} & J_n\left(\frac{nJ_n}{z}\sin\theta + iJ_n'\cos\theta\right)\frac{v_z}{v_\perp} & J_n^2\frac{v_z^2}{v_\perp^2} \end{bmatrix} \qquad (13a)$$

$$\tilde{\mathbf{R}}_n = \begin{bmatrix} 0 & J_n\left(\frac{nJ_n}{z}\cos\theta + iJ_n'\sin\theta\right)v_\perp X & 0 \\ J_n\left(\frac{nJ_n}{z}\cos\theta - iJ_n'\sin\theta\right)U & \left(\frac{2nv_z}{z}\sin\theta + v_d\right)J_n^2 X & J_n^2 W \\ 0 & J_n^2 v_z X & 0 \end{bmatrix} \qquad (13b)$$

We have made use of some identities of Bessel functions to get (12), see Ref. 1.

$\omega_{ps} = \left(\dfrac{4\pi q_s n_0^2}{m_s}\right)^{1/2}$ is the plasma frequency of species $s$. The equilibrium distribution is

normalized to unity, i.e. $\int f_0(v_z, v_\perp) d^3\mathbf{v} = 1$.

(12) is the general form of the susceptibility tensor for a plasma with a equilibrium transverse

electric field. Setting $v_d = 0$ and $\theta = 0$ in (12)(13), one obtains the result derived in previous works, as expected.

To proceed we need to specify the distribution function. We assume a Maxwellian distribution for perpendicular velocity and leave the parallel distribution unspecified, i.e.

$$f_0(v_z, v_\perp) = h(v_z) \frac{1}{\pi w_\perp^2} \exp\left(-\frac{v_\perp^2}{w_\perp^2}\right) \qquad (14)$$

The susceptibility tensor is further evaluated to be

$$\chi_s = \frac{\omega_{ps}^2}{\omega^2} \frac{2\tilde{\omega}\langle v_z\rangle}{k_z w_\perp^2}\hat{\mathbf{z}}\hat{\mathbf{z}} + \frac{\omega_{ps}^2}{\omega^2} e^{-\lambda}\left(\sum_{n=-\infty}^{+\infty} \mathbf{Y}_n(\lambda) + v_d \sum_{n=-\infty}^{+\infty} \mathbf{S}_n(\lambda)\right) \qquad (15)$$

With

$$\mathbf{Y}_n = \begin{bmatrix} \left[\dfrac{n^2 I_n}{\lambda} + 2\lambda\left(I_n - I_n'\right)\sin^2\theta\right]A_n & -\left(in + \lambda\sin 2\theta\right)\left(I_n - I_n'\right)A_n & \left[\dfrac{nI_n}{\lambda}\cos\theta - i\left(I_n - I_n'\right)\sin\theta\right]\dfrac{k_\perp}{\Omega_s}B_n \\ \left(in - \lambda\sin 2\theta\right)\left(I_n - I_n'\right)A_n & \left[\dfrac{n^2 I_n}{\lambda} + 2\lambda\left(I_n - I_n'\right)\cos^2\theta\right]A_n & \left[\dfrac{nI_n}{\lambda}\sin\theta + i\left(I_n - I_n'\right)\cos\theta\right]\dfrac{k_\perp}{\Omega_s}B_n \\ \left[\dfrac{nI_n}{\lambda}\cos\theta + i\left(I_n - I_n'\right)\sin\theta\right]\dfrac{k_\perp}{\Omega_s}B_n & \left[\dfrac{nI_n}{\lambda}\sin\theta - i\left(I_n - I_n'\right)\cos\theta\right]\dfrac{k_\perp}{\Omega_s}B_n & \dfrac{2(\tilde{\omega} - n\Omega_s)I_n}{k_z w_\perp^2}B_n \end{bmatrix}$$

(16a)

$$\mathbf{S}_n = \begin{bmatrix} 0 & \dfrac{2\Omega_s}{k_z w_\perp^2}\left[nI_n\cos\theta - i\lambda\left(I_n - I_n'\right)\sin\theta\right]A_n & 0 \\ \dfrac{2\Omega_s}{k_\perp w_\perp^2}\left[nI_n\cos\theta + i\lambda\left(I_n - I_n'\right)\sin\theta\right]A_n & \dfrac{2v_d}{w_\perp^2}\left(1 + \dfrac{2n\Omega_s}{k_z v_d^2}\sin\theta\right)(1 + A_n)I_n & \dfrac{2I_n}{w_\perp^2}\left(\langle v_z\rangle + B_n\right) \\ 0 & \dfrac{2I_n}{w_\perp^2}\left(\langle v_z\rangle + B_n\right) & 0 \end{bmatrix} \qquad (16b)$$

Where $\lambda = \dfrac{1}{2}k_\perp^2 \rho_s^2 = \dfrac{1}{2}k_\perp^2\left(\dfrac{w_\perp}{\Omega_s}\right)^2$, $\langle v_z\rangle = \int_{-\infty}^{+\infty} v_z h(v_z) dv_z$, and $I_n(\lambda) = i^{-n} J_n(i\lambda)$ is

the modified Bessel function, $A_n$ and $B_n$ are defined through integration over parallel velocity,

$$A_n = \int_{-\infty}^{+\infty} \frac{H(v_z)}{\tilde{\omega} - k_z v_z - n\Omega} dv_z \qquad (17a)$$

$$B_n = \int_{-\infty}^{+\infty} \frac{v_z H(v_z)}{\tilde{\omega} - k_z v_z - n\Omega} dv_z \qquad (17b)$$

With $H(v_z) = -\left(\tilde{\omega} - k_z v_z\right)h(v_z) + \dfrac{k_z w_\perp^2}{2}h'(v_z)$. For a given distribution of parallel motion, evaluation of $A_n$ and $B_n$ will yield the final form of the susceptibility tensor.

### III.    VERIFICATION OF THE SUSCEPTIBILITY TENSOR BY LORENTZ TRANSFORMATION

In this section we verify the susceptibility tensor derived in Sec. II through Lorentz transformation. We have derived the susceptibility tensor through the method of integration along unperturbed trajectory in the laboratory frame. In the frame drifting at velocity $v_d$ along $\hat{\mathbf{y}}$, the transverse equilibrium electric field vanishes, so the susceptibility tensor in this drifting frame is the usual form derived for the static plasmas. One obtains it by setting $v_d = 0$ in (15) and (16a,b),

$$\overline{\boldsymbol{\chi}}_s = \frac{\overline{\omega}_{ps}^2}{\overline{\omega}} \frac{2 \langle v_z \rangle}{\overline{k}_z w_\perp^2} \hat{\mathbf{z}}\hat{\mathbf{z}} + \frac{\overline{\omega}_{ps}^2}{\overline{\omega}^2} e^{-\overline{\lambda}} \sum_{n=-\infty}^{+\infty} \mathbf{Y}_n\left(\overline{\lambda}\right) \qquad (18)$$

We use an over-bar to denote the quantities in the drifting frame. To get the susceptibility tensor in the laboratory frame by Lorentz transformation from (18), one needs to write the susceptibility tensor in the covariant form. This method has been widely adopted in the beam plasma research before[11,12]. We adopt the procedure presented by Mckinstrie[11]. The current 4-vector is related to the perturbed 4-vector potential through

$$\frac{4\pi}{c} j^\mu = \chi^\mu_\nu A^\nu \qquad (19)$$

Where the Einstein summation rule is assumed. The Greeks stands for time-space indices running from 0 to 3. (19) is Lorentz covariant and as well as gauge invariant. We use a radiation gauge, i. e. setting $\phi = A^0 = 0$, to relate the spacial components of the susceptibility tensor to the conductivity tensor by

$$\overline{\chi}^i_j = \frac{4\pi\overline{\omega}}{ic^2} \overline{\sigma}^i_j \qquad (20)$$

Where Latin letters stand for spacial indices ranging from 1 to 3. Obviously, the covariant components defined in (20) are their counterparts in (18) multiplied by $\overline{\omega}^2$ except for an irrelevant constant. From the current continuity relation and the gauge invariance, we have a general relation

$$k_\mu \chi^\mu_\nu = \chi^\mu_\nu k^\nu = 0 \qquad (21)$$

Applying this relation on the susceptibility tensor in the drifting frame, one obtains

$$\overline{\chi}^0_j = \frac{c}{\overline{\omega}} \sum_{i=1}^3 \overline{k}_i \overline{\chi}^i_j \qquad (22a)$$

$$\overline{\chi}^j_0 = -\frac{c}{\overline{\omega}} \sum_{i=1}^3 \overline{k}_i \overline{\chi}^j_i \qquad (22b)$$

$$\overline{\chi}^0_0 = -\frac{c}{\overline{\omega}} \sum_{i=1}^3 \overline{k}_i \overline{\chi}^0_i = \frac{c}{\overline{\omega}} \sum_{i=1}^3 \overline{k}_i \overline{\chi}^i_0 \qquad (22c)$$

We have taken the metrics $g_{\mu\nu} = diag(1,-1,-1,-1)$. $\overline{k}_i$ is the space component of the wave vector, e. g. $\overline{k}_1 = \overline{k}_x = \overline{k}_\perp \cos\theta$, etc. The laboratory frame drifts at a velocity $-v_d \hat{\mathbf{y}}$ relative to the drifting frame, the matrix of Lorentz transformation from the drifting frame to the laboratory frame is

$$L_\nu^\mu(\beta_d) = \begin{bmatrix} \gamma & 0 & \gamma\beta_d & 0 \\ 0 & 1 & 0 & 0 \\ \gamma\beta_d & 0 & \gamma & 0 \\ 0 & 0 & 0 & 1 \end{bmatrix} \qquad (23)$$

Where $\beta_d = \dfrac{v_d}{c}$ and $\gamma = \dfrac{1}{\sqrt{1-\beta_d^{\ 2}}}$. $L_\nu^\mu(-\beta_d)$ is the inverse transformation. The transformation of the susceptibility tensor from the drifting frame to the laboratory frame is thus given by $\chi_\nu^\mu = L_\tau^\mu(\beta_d)\overline{\chi}_\sigma^\tau L_\nu^\sigma(-\beta_d)$. Writing the spacial components of the susceptibility tensor in the laboratory frame in the matrix form, we have

$$\left[\chi_j^i\right] = \begin{bmatrix} \overline{\chi}_1^1 & \gamma\left(\overline{\chi}_2^1 - \beta_d\overline{\chi}_0^1\right) & \overline{\chi}_3^1 \\ \gamma\left(\overline{\chi}_1^2 + \beta_d\overline{\chi}_1^0\right) & \gamma^2\left(\overline{\chi}_2^2 + \beta_d\overline{\chi}_2^0 - \beta_d\overline{\chi}_0^2 - \beta_d^{\ 2}\overline{\chi}_0^0\right) & \gamma\left(\overline{\chi}_3^2 + \beta_d\overline{\chi}_3^0\right) \\ \overline{\chi}_1^3 & \gamma\left(\overline{\chi}_2^3 - \beta_d\overline{\chi}_0^3\right) & \overline{\chi}_3^3 \end{bmatrix} \quad (24)$$

We first multiply each component on the right side of (18) by $\overline{\omega}^2$ to get the spacial components of the covariant susceptibility tensor. We substitute the resulting components into (22a,b,c) and (24), and then divide the results by $\omega^2$ to obtain the susceptibility tensor in the laboratory frame derived in the previous section. Taking the non-relativistic limit, i. e. setting $\gamma = 1$, $\overline{\omega}_{ps}^{\ 2} = \omega_{ps}^{\ 2}$, $\overline{\omega} = \tilde{\omega} = \omega - k_y v_d$ and $\overline{k}_i = k_i$, we obtain exactly the same form of the susceptibility tensor as that presented in (15) and (16a,b). The operation is tedious but straightforward. In the process we have used some identities of modified Bessel functions,

$$\sum_{n=-\infty}^{+\infty} nI_n = 0, \quad \sum_{n=-\infty}^{+\infty}\left(I_n - I_n'\right) = 0 \quad \text{and} \quad e^{-\lambda}\sum_{n=-\infty}^{+\infty} I_n(\lambda) = 1.$$

The approach of Lorentz transformation to derive (15) and (16a,b) is not less complicated than the method of integration along unperturbed trajectory in the laboratory frame for the present case, since we have to first derive (18) in the drift frame and then transform to the laboratory frame using (24). Nevertheless it is usually more convenient to adopt the transformation method for plasmas consisting of multiple species drifting at different velocities.

IV.    WAVES PROPAGATING PARALLEL TO THE MAGNETIC FIELD

It is very difficult to analyze the general form of the dispersion relation if a background electric is involved. In this section, we give an analysis for the electromagnetic wave propagating parallel to the background magnetic field, so $k_\perp = 0$, $\mathbf{k} = k\hat{\mathbf{z}}$, and $\tilde{\omega} = \omega$. To further simplify the problem, both the parallel and the perpendicular distributions are assumed to be Maxwellian with the same temperature, and the parallel drift velocity is set to be 0. Simple manipulation then yields

$$A_n = \frac{\omega}{kw_s} Z_p(\xi_n) \qquad (25a)$$

$$B_n = \frac{\omega}{k}\left(1 + \xi_n Z_p\right) \qquad (25b)$$

Where $Z_p$ is the well known plasma dispersion function, $\xi_n = \dfrac{\omega - n\Omega_s}{kw_s}$, and $w_s = \sqrt{2T_s/m_s}$

is the particle thermal velocity of species $s$. Since $k_\perp = 0$ leads to $\lambda = \dfrac{1}{2}k_\perp{}^2\rho_s{}^2 = 0$, only a few low order terms in the summation in (15) need to be kept. Simple operation shows that the wave is governed by the equation

$$
\begin{bmatrix}
1 - N^2 + \dfrac{1}{2}\sum_{s=i,e}\dfrac{\omega_{ps}{}^2}{\omega^2}(A_{+1} + A_{-1}) & \dfrac{i}{2}\sum_{s=i,e}\dfrac{\omega_{ps}{}^2}{\omega^2}(A_{+1} - A_{-1}) & 0 \\[2ex]
-\dfrac{i}{2}\sum_{s=i,e}\dfrac{\omega_{ps}{}^2}{\omega^2}(A_{+1} - A_{-1}) & 1 - N^2 + \sum_{s=i,e}\dfrac{\omega_{ps}{}^2}{\omega^2}\left[\dfrac{1}{2}(A_{+1} + A_{-1}) + \dfrac{2v_d{}^2}{w_s{}^2}(1 + A_0)\right] & \sum_{s=i,e}\dfrac{\omega_{ps}{}^2}{\omega^2}\dfrac{2v_d}{w_s{}^2}B_0 \\[2ex]
0 & \sum_{s=i,e}\dfrac{\omega_{ps}{}^2}{\omega^2}\dfrac{2v_d}{w_s{}^2}B_0 & 1 + \sum_{s=i,e}\dfrac{\omega_{ps}{}^2}{\omega^2}\dfrac{2\omega}{kw_s{}^2}B_0
\end{bmatrix}
\begin{bmatrix} E_x \\ E_y \\ E_z \end{bmatrix} = 0 \qquad (26)
$$

Where $N = kc/\omega$ is the diffraction index.

To proceed the analysis, we assume that the strength of the electric field is so small that the drift velocity term only contributes a small correction. Under this condition, we can make an analysis of the dispersion relation through a perturbation expansion.

We introduce $L = 1 + \sum_{s=i,e}\dfrac{\omega_{ps}{}^2}{\omega^2}\dfrac{2\omega}{kw_s{}^2}B_0$ for *Longitudinal*, and

$T = 1 - N^2 + \dfrac{1}{2}\sum_{s=i,e}\dfrac{\omega_{ps}{}^2}{\omega^2}(A_{+1} + A_{-1})$ for *transverse*, then the dispersion relation is given by

$$L\left[T^2 + T\sum_{s=i,e}\dfrac{\omega_{ps}{}^2}{\omega^2}\dfrac{2v_d{}^2}{w_s{}^2}(1 + A_0) - \dfrac{1}{4}\left(\sum_{s=i,e}\dfrac{\omega_{ps}{}^2}{\omega^2}(A_{+1} - A_{-1})\right)^2\right] = T\left(\sum_{s=i,e}\dfrac{\omega_{ps}{}^2}{\omega^2}\dfrac{2v_d}{w_s{}^2}B_0\right)^2$$

$$(27)$$

There are two independent transverse electromagnetic modes and one longitudinal electrostatic mode if the transverse electric field is absent. With a transverse equilibrium electric field present, the three modes are coupled and their polarizations are also modified. We give an analysis for the transverse modes using perturbation method for small drift velocity. Multiplying (27) by $L^{-1}$ and introducing a small parameter $\sigma = \sum\limits_{s=i,e} \dfrac{\omega_{ps}^2}{\omega^2} \dfrac{2v_d^2}{w_s^2}\left(1+A_0\right) - L^{-1}\left(\sum\limits_{s=i,e} \dfrac{\omega_{ps}^2}{\omega^2} \dfrac{2v_d^2}{w_s^2} B_0\right)^2$, one

obtains the solution to (27) approximately

$$T = \pm\frac{1}{2}\sum_{s=i,e}\frac{\omega_{ps}^2}{\omega^2}\left(A_{+1}-A_{-1}\right)-\frac{1}{2}\sigma \qquad (28)$$

Using the definition of symbols, we get the two branches of waves

$$N^2 = 1 + \sum_{s=i,e}\frac{\omega_{ps}^2}{\omega^2}\frac{\omega}{kw_s}Z_p\left(\frac{\omega\pm\Omega_s}{kw_s}\right)+\frac{1}{2}\sigma \qquad (29)$$

Substitution of (28) in (26) gives the ratio of the transverse electric field components

$$\frac{E_x}{E_y} = \frac{i\sum\limits_{s=i,e}\dfrac{\omega_{ps}^2}{\omega^2}\left(A_{+1}-A_{-1}\right)}{\pm\sum\limits_{s=i,e}\dfrac{\omega_{ps}^2}{\omega^2}\left(A_{+1}-A_{-1}\right)-\sigma} \qquad (30)$$

Obviously the two circularly polarized electromagnetic waves become elliptically polarized due to the presence of the transverse equilibrium electric field. On the other hand, from (26) we notice that a longitudinal component of the perturbed electric field is present and proportional to the equilibrium electric field.

We present an approximate result in the cold plasma limit. The plasma consists of hydrogen ions and electrons. Using the asymptotic expansion of the plasma dispersion function and keeping the leading terms one obtains the reduced form of (29)

$$N^2 = 1 - \sum_{s=i,e}\frac{\omega_{ps}^2}{\omega}\frac{1}{\omega\pm\Omega_s}+\frac{1}{2}\sigma \qquad (31)$$

and

$$\sigma = -N^2\frac{\omega_p^2}{\omega^2}\frac{v_d^2}{c^2}\left(1+L^{-1}\frac{\omega_p^2}{\omega^2}\right)^2 \qquad (32)$$

Where $\omega_p^2 = \omega_{pi}^2+\omega_{pe}^2$. Usually in magnetically confined plasmas, the drift velocity is in the order of the ion thermal velocity. Normalizing the velocity by ion thermal velocity, one obtains from (31)

$$\left(1+\delta\right)N^2 = 1 - \frac{\omega_p^2}{\left(\omega\pm\Omega_e\right)\left(\omega\pm\Omega_i\right)} \qquad (33)$$

with

$$\delta = \frac{1}{2}\left(\frac{v_d}{w_i}\right)^2\frac{m_i\beta_i}{m_e}\frac{\Omega_i^2}{\omega^2-\omega_p^2} \qquad (34)$$

Where $\beta_i = \dfrac{8\pi n_i T_i}{B_0^2}$ the ratio between the ion thermal pressure and the magnetic pressure.

From (33) (34) one notes that the equilibrium electric field has significant contributions only if the mode frequency is very close to the plasma Langmuir frequency. We make a simple estimate for a typical tokamak plasma[13]. For a hydrogen plasma, setting $\beta_i = 0.04$, the drift velocity

$$\frac{v_d}{w_i} = 1 \text{ and ion temperature } T_i = 10 \text{ KeV, noting that } \frac{\omega_p}{\Omega_i} \simeq \left( \frac{m_i}{m_e} \frac{c^2 \beta_i}{w_i^2} \right)^{1/2} \simeq 1860 \text{ one finds}$$

that the $N^2(\omega)$ relation is significantly modified by the drift velocity if

$| \omega - \omega_p | / \Omega_i \sim O(0.01)$. However, in solving (27) we have assumed that $\delta$ is small and used a perturbation method. Then if the wave frequency approaches $\omega_p$ to make $\delta$ of order unity, the solving process is not valid any more. From (33) and (34), we observe a resonance near $\omega_p$ and a cut-off at $\omega = \omega_p$. The cut-off is artificially introduced since one multiplies (27) by $L^1$ which is approaching infinity for $\omega \sim \omega_p$. But a resonance at $\omega = \omega_p$ is indeed in existence. Starting from Eq. (27), If the wave frequency approaches the Langmuir frequency, using the adiabatic expansion of the plasma dispersion function one gets $L \sim 0$. Then there are two solutions for $T$, i. e. $T \sim 0$ and $T \sim \infty$. The latter one means a resonance, i. e. $N^2 \sim \infty$.

## V.    SUMMARY AND DISCUSSION

In this work, we have derived the susceptibility tensor of a uniform magnetized plasma when a transverse equilibrium electric field exists. The method of integration along unperturbed trajectory is adopted to get the perturbed distribution function. The general form of the tensor is given by Eqs. (12) and (13a,b). Assuming the perpendicular motion of all species obey the Maxwellian distribution with a common shift velocity, we obtain the susceptibility tensor as given in (15) (16a,b). Then we adopt the Lorentz transformation to verify the derivations in the laboratory frame.

The analysis for the general electromagnetic waves in a magnetized plasma is almost intractable. As an example of the application, we present the analysis of the electromagnetic wave propagating parallel to the background magnetic field in the cold plasma limit for a simple plasma consisting of isotropic electrons and hydrogen ions with the same temperature. Even for such a simple case, full analytical treatment is still very difficult. The otherwise independent two transverse electromagnetic waves and one longitudinal oscillation are coupled through the drift velocity. The wave polarizations are all modified due to the coupling effect. The coupling also induces a resonance near the plasma Langmuir frequency. An approximate analysis for small drift velocities reveals that the $N^2(\omega)$ is significantly changed for the wave frequency close to $\omega_p$.

Changed $N^2(\omega)$ relation means that the existence of the equilibrium electric field will alternate the Faraday rotation, and consequently influence the diagnostics such as POINT (Polarimeter Interferometer) based on the Faraday rotation[14].

For the waves propagating parallel to the background magnetic field, the effect from the equilibrium electric field is almost negligible if the wave frequencies are not so close to $\omega_p$. A qualitative change caused by the electric field is the existence of a resonance at $\omega = \omega_p$.

Usually in laboratory plasmas the $\mathbf{E} \times \mathbf{B}$ drift velocity is in the order of the ion thermal velocity. For waves travelling at arbitrary angles, it is likely that the lower frequency modes, such as the ion Bernstein wave[15] and kinetic Alfvenic waves[16], will be more influenced by the electric field than those higher frequency waves, e. g. electron Bernstein waves[17]. These are left for future investigations.